\documentclass{article}
\usepackage[utf8]{inputenc}

\usepackage{amsmath}

\usepackage{amssymb}

\newtheorem{theorem}{Theorem}
\newtheorem{corollary}[theorem]{Corollary}

\newtheorem{lemma}[theorem]{Lemma}

\newtheorem{definition}[theorem]{Definition}
\newtheorem{proposition}[theorem]{Proposition}

\newtheorem{remark}[theorem]{Remark}

\newenvironment{proof}{\noindent\bf{Proof.}\rm}{\hfill$\blacksquare$\bigskip}

\newcommand{\items}{\mathcal{M}}

\begin{document}

\title{From multi-allocations to allocations, \\ with subadditive valuations}

\author{Uriel Feige}

\maketitle

\begin{abstract}
  We consider the problem of fair allocation of $m$ indivisible items to $n$ agents with monotone subadditive valuations. For integer $d \ge 2$, a $d$-multi-allocation is an allocation in which each item is allocated to at most $d$ different agents. We show that $d$-multi-allocations can be transformed into allocations, while not losing much more than a factor of $d$ in the value that each agent receives. 
  One consequence of this result is that for 
  allocation instances with equal entitlements and subadditive valuations, 
  if $\rho$-MMS $d$-multi-allocations exist, then so do $\frac{\rho}{4d}$-MMS allocations. Combined with recent results of Seddighin and Seddighin [EC 2025], this implies the existence of $\Omega(\frac{1}{\log\log n})$-MMS allocations.
\end{abstract}

\section{Introduction}

We consider allocations of a set $\items$ of $m$ indivisible items to $n$ agents. Each agent $i$ has a valuation function $v_i$ over the set of items, that is normalized ($v_i(\emptyset) = 0$), monotone ($v_i(S) \le v_i(T)$ for $S \subset T \subseteq \items$), and subadditive ($v_i(S) + v_i(T) \ge v_i(S \cup T)$). 
An allocation $A = A_1, \ldots, A_n$ is a collection of $n$ disjoint subsets of $\items$, giving bundle $A_i$ (we use the terms {\em bundle} and {\em set} interchangeably) to agent $i$ (for every $1 \le i \le n$). A multi-allocation is a collection of $n$ possibly intersecting subsets $A_1, \ldots, A_n$ of $\items$, giving bundle $A_i$ to agent $i$. For a positive integer $d$, a $d$-multi-allocation is one in which every item is a member of at most $d$ of the sets. If $d$ is not an integer (this may happen for example if  we select $d = \log n$), we interpret the term $d$-multi-allocation as meaning $\lceil d \rceil$-multi-allocation (rounding $d$ up to the nearest integer). 

The main question that we study in this work is as follows. Suppose that we are given as input a $d$-multi-allocation $A = A_1, \ldots, A_n$, and we wish to output an allocation $A' = A'_1, \ldots, A'_n$. How much of the initial value $v_i(A_i)$ that agents had in $A$ can be preserved in $A'$? We will only consider allocations in which $A'_i \subseteq A_i$ for every $i$, as it might be that an agent $i$ has no value for items outside her initial set $A_i$. Our main theorem is the following.

\begin{theorem}
    \label{thm:main}
    Let $d \ge 2$ be a perfect power of~2, and let $A = A_1, \ldots, A_n$ be a $d$-multi-allocation. 
    Suppose that every agent $i$ has a subadditive valuation $v_i$ with $\max_{e\in A_i} v_i(e) = \delta_i$. 
    Then there exists an allocation $A' = A'_1, \ldots, A'_n$ in which 
    $$v_i(A'_i) \ge \frac{v_i(A_i)}{d} - \frac{d-1}{d}\delta_i$$
    for every agent $i$.
\end{theorem}

In the special case that the valuations are additive ($v_i(S) = \sum_{e\in S} v_i(e)$ for every $S \subset \items$), Theorem~\ref{thm:main} is known to hold even if $d$ is not a power of~2. 
(This follows from standard rounding techniques for linear programs. For a survey of these techniques and additional properties that they have, see for example~\cite{BEF22BoBW}, where the technique is referred to as {\em faithful implementations of fractional allocations}.) 
An example in which no allocation offers a better guarantee than that of the theorem is when there are $d$ agents and $d-1$ items, each of value~1. For additive valuations, this implies that $\delta_i = 1$. In the $d$-multi-allocation that gives every item to all agents, every agent gets value $d-1$ (hence $v_i(A_i) = d-1$). However, in every allocation, some agent gets value $0 = \frac{v_i(A_i)}{d} - \frac{d-1}{d}\delta_i$.

\subsection{Applications to fair allocations}

In a setting with a set $\items$ of $m$ items and $n$ agents, the maximin share (MMS) of an agent $i$ is the maximum value that $i$ can guarantee to herself as a ``cutter" in a cut and choose game. In this game, $i$ first partitions $\items$ into $n$ disjoint bundles. Then, each other agent selects an arbitrary bundle of her choice, and $i$ receives the bundle that remains. Formally,
$$MMS(\items, v_i, n) = \max_{P_1, \ldots, P_n \in \mathcal{P}} \min_j v_i(P_j)$$
where $\mathcal{P}$ denotes the set of all partitions of $\items$.

An allocation is an MMS allocation if every agent receives a bundle of value at least as high as her MMS. For settings in which agents have equal entitlements, MMS is one of the prominent notions of fairness. Unfortunately, for many classes of valuation functions, including subadditive, MMS allocations need not exist. Consequently, one considers also $\rho$-MMS allocations, in which each agent gets at least a $\rho$-fraction of her MMS. It is an active research direction to determine the largest value of $\rho$ that can always be guaranteed, for various classes of valuation functions. (See Section~\ref{sec:related} for more details.) Let $\rho_{SA}$ denote the corresponding value of $\rho$ when valuations are subadditive. The exact value of $\rho_{SA}$ is not known, and it is not even known whether $\rho_{SA}$ can be lower bounded by a constant, or tends to~0 as $n$ grows. 

There are relatively simple examples showing that $\rho_{SA} \le \frac{1}{2}$~\cite{GhodsiHSSY22}. This holds even if there are only two agents and four items. There are special cases in which it is known that $\rho_{SA} = \frac{1}{2}$. They include the case that $n \le 4$, and the case when agents can be partitioned into two groups, where all agents within the same group have the same valuation function~\cite{cCMS25}. A recent result~\cite{CM25} shows that this also holds for so called {\em graphical} allocation instances, in which each item has positive value to at most two agents. Such instances can be represented as graphs (or rather multi-graphs, we allow parallel edges), in which agents are vertices and each item is an edge joining two agents, indicating that other agents have 0 value for the item. 

A by-product of the proof of
Theorem~\ref{thm:main} 
is a strengthening of this last result, replacing $MMS(\items, v_i, n)$ by the potentially much higher value of $MMS(\items, v_i, 2)$.

    \begin{corollary}
        \label{thm:graphical}
    Every multi-graphical allocation instance with subadditive valuations always has allocations in which every agent $i$ gets a bundle of value at least $\frac{1}{2} \cdot MMS(\items, v_i, 2)$.
    \end{corollary}

We remark that in~\cite{CM25}, a result similar to Corollary~\ref{cor:graphical} was proved, but with subadditive valuations replaced by XOS valuations (XOS is a strict subclass of subadditive). 

Going beyond graphical instances to geneal instances, Theorem~\ref{thm:main} allows us to formulate a lemma that serves as a convenient tool for getting improved lower bounds on the value of $\rho_{SA}$. We refer to this lemma as the {\em subadditive MMS sampling lemma}. In stating the lemma, we use the following notation. For $d \ge 1$ 
we denote the smallest power of~2 that is at least as large as $d$ by $\hat{d}$. That is, $\hat{d} = 2^{\lceil \log d \rceil}$.

\begin{lemma} [Subadditive MMS sampling lemma]
\label{lem:sampling}
    Suppose that there is a positive integer $d \ge 2$ and a constant $\rho > 0$ such that every allocation instance with subadditive valuations has a $\rho$-MMS $d$-multi-allocation. Then $\rho_{SA} \ge \rho / (2\hat{d} - 1)$.  
\end{lemma}

\begin{remark}
\label{rem:sampling}
    Lemma~\ref{lem:sampling} applies also if $d$ is not a fixed integer, but a function of $n$, provided that the function is non-decreasing in $n$. In this context, note that if $d$ is not an integer then $\hat{d} = \widehat{\lceil d \rceil}$, and so the lower bound on $\rho_{SA}$ holds as stated also if $d$ is not an integer.

    The conclusion of Lemma~\ref{lem:sampling} holds even if $\rho$-MMS $d$-multi-allocation are only guaranteed to exist conditioned on $\delta_i \le \rho \cdot v_i(\items) / (2\hat{d} - 1)$ holding for all $i$ (for $\delta_i$ as defined in Theorem~\ref{thm:main}). 
\end{remark}

We provide simple examples for how Lemma~\ref{lem:sampling} can easily be used in order to improve over known lower bounds on $\rho_{SA}$. We start with the special case of $n=8$.

    \begin{proposition}
        \label{pro:n8}
        If $n \le 8$ and agents have subadditive valuations, then a $\frac{1}{6}$-MMS allocation exists.
    \end{proposition}

    \begin{proof}
    Break the set of eight agents into two groups, each with four agents. For each group, there is a $\frac{1}{2}$-MMS allocation~\cite{cCMS25}. The combination of the allocations for the two groups is a $\frac{1}{2}$-MMS 2-multi-allocation. By Lemma~\ref{lem:sampling}, this implies the existence of a  $\frac{1}{6}$-MMS allocation.     
    \end{proof}



    

We now turn to general values of $n$. Recently it was shown that $\Omega(\frac{1}{(\log\log n})^2)$-MMS allocations exist for subadditive valuations~\cite{SS2025}. An intermediate step of the proof established that $\frac{1}{2}$-MMS $\Omega(\frac{1}{(\log\log n}))$-multi-allocations exist for subadditive valuations. Combining this intermediate result with our Lemma~\ref{lem:sampling}, we immediately obtain the following corollary.

\begin{corollary}
    \label{cor:loglogn}
    With subadditive valuations, $\rho_{SA} \ge \Omega(\frac{1}{\log\log n}$).
\end{corollary}

The hidden constant in the $\Omega$ notation is not less that $\frac{1}{8}$. (We consider here $n \ge 4$, so that $\log\log n \ge 1$.)
{See more details in Section~\ref{sec:loglogn}.}




\subsection{Related work}
\label{sec:related}

The trigger to our work comes from a line of work initiated by~\cite{SS24}. They used a two-step framework in order to design $\rho$-MMS allocations for subadditive valuations. In its cleanest form, the framework is as follows.

\begin{enumerate}
    \item For some parameter $d$, design a $\frac{1}{2}$-MMS $d$-multi-allocation. 
    \item Derive from this multi-allocation a $\frac{1}{2 f(d)}$-MMS allocation, for some function $f(d)$.
\end{enumerate}  

Results using this approach are stronger the smaller $d$ is, and if the function $f(d)$ does not grow too quickly. To get values of $d$ that are sublinear in $n$, one may use a result of~\cite{Feige09} that implies that there always is an allocation that gives at least $\frac{1}{2}$-MMS to at least half the agents. From this one can obtain in a straightforward way $d \le \log n$, 
To bound $f(d)$, it is proposed in~\cite{SS24} to use uniformly random sampling (each item is given independently at random to one of the agents that holds a copy of it in the multi-allocation). It is quite easy to show that in expectation, each agent maintains $\frac{1}{d}$ of her value, suggesting that $f(d) = O(d)$. However, it is much more difficult to show that with positive probability, it holds simultaneously for all agents that each of them gets a value not much lower than the expectation. For this one uses concentration results for submodular function (originating in~\cite{schechtman2003concentration}). However, using these concentration results is not easy. For example, step~1 above was modified in~\cite{SS24} so as to require additional properties of the multi-allocation. Moreover, despite having $d = O(\log n)$, in~\cite{SS24} they were only able to show the existence of $\Omega(\frac{1}{\log n \log\log n})$-MMS allocations, and with a very poor leading constant hidden by the $\Omega$ notation. With some modifications to step~1 and improved analysis for step~2 (in particular, deriving concentration bounds that are stronger than those taken from~\cite{schechtman2003concentration}), this bound was improved to~$\frac{1}{14 \log n}$-MMS in~\cite{FH25}.


Lemma~\ref{lem:sampling} provides a simpler way of going from multi-allocations to allocations. There is no need for the multi-allocation of step~1 to have any additional properties. The associated function $f(d)$ satisfies $f(d) = O(d)$, which is best possible. Moreover, the hidden constant in the $O$ notation is not more than~4, avoiding larger constants that often appear in analysis based on concentration of submodular functions. For example, our Lemma~\ref{lem:sampling} can be used to replace  the $\frac{1}{14 \log n}$-MMS allocations of~\cite{FH25} by $\frac{1}{8 \log n}$-MMS allocations. More dramatically, saving both a $\log\log n$ factor and a huge constant, Lemma~\ref{lem:sampling} can be used to replace  the recent $\frac{1}{432000 (\log\log n)^2}$-MMS allocations of~\cite{SS2025} {by $\frac{1}{8 \log\log n}$-MMS allocations} (Corollary~\ref{cor:loglogn}).
Perhaps most importantly, the relation $f(d) = O(d)$ holds for all $d$, including constant $d$, whereas for the random sampling approach the relation $f(d) = O(d)$ requires $d$ to grow with $n$. In particular, Lemma~\ref{lem:sampling} implies that to prove existence of $\Omega(1)$-MMS allocations, it suffices to prove existence of $\Omega(1)$-MMS $d$-multi-allocations, for some constant $d$. Such an implication was not previously known.

Other works lower bounded $\rho_{SA}$ as a function of $m$ (the number of items) rather than as a function of $n$ (number of agents). Observe that we may assume that $n \le m$, as otherwise the MMS is~0, and every allocation is an MMS allocation. In~\cite{GhodsiHSSY22} it was shown that $\rho_{SA} \ge \frac{1}{10 \log m}$. The case in which agents have arbitrary (not necessarily equal) entitlements was studied in~\cite{FG25}. The results there, when translated to the equal entitlement setting, imply $\frac{1}{k+1}$-MMS allocations for the smallest $k$ satisfying $m \le k^k$. 

Approximate MMS allocations were also studied for subclasses of subadditive valuations. 

\begin{itemize}
    \item For additive ($A$) valuations we have $\frac{1}{4} + \frac{3}{3836} \le \rho_A \le \frac{39}{40}$~\cite{akrami2023breaking, FST21}.
    \item For submodular ($SM$) valuations we have $\frac{10}{27} \le \rho_{SM} \le \frac{2}{3}$~\cite{BUF23, KKM23}.
    \item For XOS (XOS) valuations we have $\frac{4}{17} \le \rho_{XOS} \le \frac{1}{2}$~\cite{FG25, GhodsiHSSY22}.
\end{itemize}

Our Corollary~\ref{thm:graphical} concerns MMS allocations for allocation instances that can be represented as multi-graphs. Graphical instances (without parallel edges) were studied in~\cite{CFKS23}, where it was shown that EFX allocations exist for every class of monotone valuations, but might require allocating items to agents that have value~0 for them (allocating edges to non-endpoints of the edge). For the case of additive valuations, this extends to bipartite multi-graphs~\cite{ADKMR24}. Approximate MMS allocations for multi-graphical instances were recently studied in~\cite{CM25}, where among other results it was shown that MMS-allocations exist for additive valuations, $\frac{2}{3}$-MMS allocations exist for XOS valuations, and $\frac{1}{2}$-MMS allocations exist for subadditive valuations. Our Corollary~\ref{thm:graphical} is stronger than this last result, as it replaces $MMS(\items, v_i, n)$ by $MMS(\items, v_i, 2)$.

Multi-allocations as a form of ``resource augmentation" are studied in~\cite{AEFFG25}. For example, it is shown that with additive valuations, there always are MMS 2-multi-allocations, in which at most $n-2$ items have copies.




\section{Proofs}

Our proof of Theorem~\ref{thm:main} has three parts. In Section~\ref{sec:picking} we analyze an allocation game with two agents, and prove that each agent $i$ can ensure a value of at least $\frac{1}{2}(v_i(\items) - \delta_i)$. In section~\ref{sec:graphical} we extend this game to graphical instances with any number of agents, and show that also in this case, each agent $i$ can ensure a value of at least $\frac{1}{2}(v_i(\items) - \delta_i)$. This implies Theorem~\ref{thm:main} in the special case of $d=2$. In Section~\ref{sec:induction} we use induction on $k$ to prove the theorem for $d = 2^k$, for every $k \ge 1$. Lemma~\ref{lem:sampling} is then proved in Section~\ref{sec:lemma}. In Section~\ref{sec:loglogn} we present more details about the proof of Corollary~\ref{cor:loglogn}, providing bounds on the constant hidden under the $\Omega$ notation.

\subsection{Alternating picking sequences}
\label{sec:picking}

We consider the following game, involving two players, $p$ and $q$, and a set $\items$ of $m$ items. Player $p$ has a monotone valuation function $v$ over the items. 
A picking sequence $S = s_1, \ldots, s_m \in \{p,q\}^m$ determines the order in which players can pick items, where the player to pick an item in round $r$ (for $1 \le r \le m$) is $s_r$. Let $T$ denote the set of items picked by player $p$. The payoff to player $p$ is $v(T)$, and for player $q$ it is $-v(T)$. Hence the game is a 0-sum finite game. Being a game with full information, the maximin/minimax strategies for this game are deterministic. The maximin value of this game for player $p$ is denoted by $\omega(S, \items, v)$. 

$S$ is said to be alternating (or Round Robin) if it starts with either $p$ or $q$, and then in every round switches the player whose turn it is to pick. Hence, there are two possible alternating sequences, one starting with $p$ (which we call $S_p$), the other by $q$ (which we call $S_q$). We wish to lower bound $\omega(S, \items, v)$ over all subadditive valuations $v$ and alternating picking sequences $S$. For future reference, we summarize the relevant notation in the following definition.

\begin{definition}
    \label{def:game}
    In the game described above, $\omega(S_p, \items, v)$ denotes the maximin value that agent $p$ (with valuation $v$) can guarantee to herself in the alternating picking sequence $S_p$ in which $p$ picks first, and $\omega(S_q, \items, v)$ denotes the maximin value that agent $p$ can guarantee to herself in the alternating picking sequence $S_q$ in which $q$ picks first.
\end{definition}

\begin{proposition}
\label{pro:greater}
    For every $v$ it holds that $\omega(S_p, \items, v) \ge \omega(S_q, \items, v)$.
\end{proposition}

\begin{proof}
    As the game is finite and of full information, the maximin strategies of the players are deterministic. Consider the picking sequence $S_q$, and let $P$ denote the maximin strategy for $p$ in this game, leading to the value $\omega(S_q, \items, v)$.

    Consider now the picking sequence $S_p$. We shall design a strategy $P'$ for $p$ that guarantees for $p$ a payoff of at least  $\omega(S_q, \items, v)$, thus establishing that $\omega(S_p, \items, v) \ge \omega(S_q, \items, v)$.

    In $S_p$, player $p$ picks first. In $P'$, we fix an arbitrary item $e_1$, and $p$ picks $e_1$ in the first round. We refer to this as a {\em one-sided pick} because this pick only limits the choices available to $q$ (who cannot pick $e_1$, because it is already taken), but we do not view $e_1$ as contributing any value to $p$. That is, for a set $T$ of additional items that $p$ picks, her total value will be $v(T)$, and not $v(T \cup \{e_1\})$.
    
    Thereafter, $p$ simulates the strategy $P$. That is, for every item that $q$ picks, $p$ replies by picking the item that $P$ dictates, where $P$ is run as if $p$ does not hold item $e_1$.

    Now there are two options.  One option is that it never happens in this run that $P$ dictates that $p$ should pick $e_1$. 
    If $m$ is even, then $p$ gets a bundle that it would get also in the picking sequence $S_q$. (Now $e_1$ is considered part of this bundle. One may think of the one-sided pick of $e_1$ as if it never happened, and instead $e_1$ was the last item that remained in the execution under $S_q$, and then $p$ who picks last in $S_q$ indeed gets $e_1$.) If $m$ is odd, then $p$ gets a bundle that it would get also in the picking sequence $S_q$, plus an additional item, $e_1$.
    Hence, regardless of whether $m$ is even or odd, $p$ gets a bundle of value at least $\omega(S_q, \items, v)$. 

    The other option is that in some round $r_1$, $P$ dictates that $p$ picks $e_1$. Player $p$ cannot do so, because she already holds $e_1$. But then, we can think of the event of picking $e_1$ not as happening in the first round, but rather at round $r_1$. That is, we undo the one-sided pick of $e_1$, and replace it by a true pick of $e_1$ in round $r_1$. So altogether, it is as if strategy $P$ was played for the first $r_1-1$ rounds (round~2 up to round $r$, because round~1 was undone). The bundles that $p$ and $q$ hold are thus legitimate outcomes of running strategy $P$ for the sequence $S_q$ for $r_1-1$ rounds. Now strategy $P'$ selects some arbitrary yet unpicked item, call it $e_2$, and dictates that $p$ picks $e_2$. Again, this pick is treated as a one-sided pick.

    Also with respect to $e_2$, there are two options, as above. In the first of these options, the game ends with $p$ holding a bundle that she could get using strategy $P$ in the picking sequence $S_q$ (or such a bundle together with the extra item $e_2$, if $m$ is odd), and hence of value at least $\omega(S_q, \items, v)$. In the second option, $P$ would dictate picking $e_2$ in some round $r_2$. But then again, we will have that the bundles that $p$ and $q$ hold are legitimate outcomes of running strategy $P$ for the sequence $S_q$ for $r_2-1$ rounds. Thus we can continue with player $p$ picking a yet unpicked item $e_3$, and continuing as above. As $|\items|$ is finite, eventually the first option will hold, establishing that $\omega(S_p, \items, v) \ge \omega(S_q, \items, v)$.
\end{proof}

\begin{remark}
We alert the reader that even though Proposition~\ref{pro:greater} holds both for additive and for subadditive valuations, some other properties that hold in the additive case fail to hold in the subadditive case. (These properties fail already when $v$ is XOS, which is a subclass of additive valuations.)

    Let $P$ and $Q$ denote that bundles that $p$ and $q$ receive when the players play their maximin strategies. In the subadditive case, it need not be that $v(P) \ge v(Q)$. For example, with $m=4$, it could be that every pair of items has value $3$, except that $v(\{e_2,e_4\}) = 5$, $v(\{e_1,e_2\}) = v(\{e_1,e_3\}) = 4$. Then under the maximin strategies, the sequence of picks might be $e_1, e_2, e_3, e_4$, with $5 = v(Q) > v(P) = 4$.

    Also, for two picking sequences, $S_1$ and $S_2$, suppose that for every $1 \le t \le n$, the length $t$ prefix of $S_2$ has at least as many picks by $p$ as the length $t$ prefix of $S_1$. In the subadditive case, this does not imply that $\omega(S_2, \items, v) \ge \omega(S_1, \items, v)$. For example, let $S_1 = qppq$ and $S_2 = pqpq$. Consider the subadditive valuation in which $v(T) = \max[\mid T \cap \{e_1,e_2\}\mid , \mid T \cap \{e_3,e_4\}\mid ]$. For this $v$, $1 = \omega(S_2, \items, v) < \omega(S_1, \items, v) = 2$.
\end{remark}

\begin{proposition}
\label{pro:sum}
    If $v$ is subadditive, then $\omega(S_p, \items, v) + \omega(S_q, \items, v) \ge v(\items)$.
\end{proposition}

\begin{proof}
    Consider the valuation function $\bar{v}$ that for every set $T$ has value $\bar{v}(T) = v(\items) - v(\items \setminus T)$. (Note that $\bar{v}$ need not be subadditive, but this does not affect our proof.) Observe that if $q$ gets $T$, then $p$ gets $\items \setminus T$.  Hence by maximizing $\bar{v}$ for herself, agent $q$ minimizes $v$ for $p$. For the picking sequence $S_p$, if $p$ plays her maximin strategy for $v$ and $q$ plays her maximin strategy for $\bar{v}$, then being a constant sum game (in every outcome of the game, the sum of payoffs is exactly $v(\items)$), $p$ gets $\omega(S_p, \items, v)$, whereas $q$ (being the second player to pick) gets $\omega(S_q, \items, \bar{v})$. Hence $\omega(S_p, \items, v) + \omega(S_q, \items, \bar{v}) = v(\items)$.

    To complete the proof, observe that subadditivity of $v$ implies that for every set $T$, $v(T) \ge v(\items) - v(\items \setminus T) = \bar{v}(T)$. Hence, $\omega(S_q, \items, v) \ge \omega(S_q, \items, \bar{v})$.
\end{proof}

\begin{corollary}
\label{cor:subadditive}
    If $v$ is subadditive and no single item has value larger than $\delta$, then $\omega(S_p, \items, v) \ge \frac{1}{2} v(\items)$,  and $\omega(S_q, \items, v) \ge \frac{v(\items) - \delta}{2}$. Moreover, $\omega(S_q, \items, v) \ge MMS(\items, v_i, 2)$. (The other inequality, $\omega(S_p, \items, v) \ge MMS(\items, v_i, 2)$, follows immediately from $\omega(S_p, \items, v) \ge \frac{1}{2} v(\items)$).
\end{corollary}

\begin{proof}
    By Proposition~\ref{pro:sum}, we have that $\omega(S_p, \items, v) + \omega(S_q, \items, v) \ge v(\items)$. By Proposition~\ref{pro:greater}, $\omega(S_p, \items, v) \ge \omega(S_q, \items, v)$. The combination of these two inequalities implies that $\omega(S_p, \items, v) \ge \frac{1}{2} v(\items)$.

    To lower bound $\omega(S_q, \items, v)$, let $\items'$ be the set of $m-1$ items that remain after $q$ picks her first item in the picking sequence $S_q$. For $\items'$, player $p$ is the one who picks first. Hence:

    $$\omega(S_q, \items, v) = \omega(S_p, \items', v) \ge \frac{1}{2} v(\items') \ge \frac{v(\items) - \delta}{2}$$
    The last inequality follows by subadditivity of $v$, and the fact that the first item taken by $q$ has value at most $\delta$.
    Note that $v(\items') \ge MMS(\items, v_i, 2)$ (because $\items'$ necessarily contains one of the bundles of the $MMS(\items, v_i, 2)$ partition), and so $\omega(S_q, \items, v) \ge MMS(\items, v_i, 2)$.
\end{proof}

\begin{remark}
    \label{rem:delta}
    Inspecting the proof of Corollary~\ref{cor:subadditive}, one can replace the interpretation of the parameter $\delta$ to be $\max_{e \in \items}[v(\items) - v(\items \setminus \{e\}]$, instead of $\max_{e \in \items}[v(e)]$. (This was implicitly done in the proof of the ``moreover" part of the corollary.) For additive valuations, this replacement changes nothing, but for subadditive valuations, this often leads to stronger lower bounds on $\omega(S_q, \items, v)$.
\end{remark}


\subsection{Multi-graph allocation instances}
\label{sec:graphical}

We consider allocation instances in which each item is desired by at most two agents, and has 0-marginal value for all other agents. Such an instance can be represented as a multi-graph $G$ ($G$ may have parallel edges). The $n$ vertices of $G$ correspond to the $n$ agents. The $m$ edges of $G$ correspond to the $m$ items. For an item $e$, all agents other then the two endpoints of the corresponding edge have~0 marginal value for the item. We refer to the set of items corresponding to the edges incident with the vertex of $i$ as $\items^{i}$.
Every agent $i$ has a 
valuation function $v_i$ over $\items^{i}$ (and 0 marginal value to each other item). 

We consider only allocations that are represented by orientations of the edges, where if an edge is oriented from $i$ to $j$, this means that agent $j$ receives the item.

For the following theorem, recall Definition~\ref{def:game}.

\begin{theorem}
    \label{thm:multigraph}
    In allocation instances represented by multi-graphs as described above, there is an allocation in which every agent $i$ receives a bundle of value at least $\omega(S_q,\items^{i},v_i)$.
\end{theorem}

\begin{proof}
    Without loss of generality we may assume that all vertices of $G$ have even degrees. (We may enforce this by adding ``auxiliary" edges between pairs of vertices of odd degree, where the auxiliary edges correspond to items that have 0 marginal value to all agents.)

    We consider the following allocation game among the $n$ agents. The game is played on the multi-graph $G$ that represents the allocation instance. There is a token that specifies whose turn it is to move. In every round, the token either {\em walks} or {\em jumps}, as explained below. Initially, the token jumps to an arbitrary vertex, say, vertex~1.  In a given round, we do the following:
    
    \begin{itemize}
        \item If all items incident with the vertex holding the token are already allocated, the token {\em jumps} to an arbitrary vertex that is incident with a yet unallocated item.
        \item Else, the agent on whose vertex the token is placed gets to pick a yet unallocated item of her choice (among those items incident with its vertex), and the token {\em walks} to the other endpoint of the corresponding edge.
    \end{itemize}
    
    The allocation game ends when all items are allocated.

    The play of the game can be decomposed into phases. A phase begins when the token jumps to the initial vertex of the phase, say, vertex $u$. Thereafter, the token walks along distinct edges of $G$ (corresponding to the items chosen by the players). The token may visit the same vertex several times during a phase. The phase ends when the token is at a vertex with no incident unallocated item. As all degrees are even, this last vertex must be the initial vertex $u$.

    A maximin strategy for an agent $i$ is a strategy that has the highest guaranteed value, where the guaranteed value of a strategy is the value under $v_i$ of the worst possible bundle that agent $i$ can get, under all possible strategies for the other agents.

    For an agent $i$, we do a case analysis.

    \begin{enumerate}
        \item The token first visited $i$ by jumping into it. In this case, from the point of view of $i$, she is beginning a phase that corresponds to a setting in which she is player $p$ in the picking sequence $S_p$. Hence, she can guarantee a value of $\omega(S_p, \items^i, v_i)$. By Proposition~\ref{pro:greater}, this is at least $\omega(S_q, \items^i, v_i)$.
        \item The token first visited $i$ by walking into it.  In this case, from the point of view of $i$, she is player $p$ in the picking sequence $S_q$, and can start playing a strategy that guarantees a value of $\omega(S_q, \items^i, v_i)$. Now there are two options.

        \begin{itemize}
            \item In future rounds, the token only enters $i$ by walking into it. In this case, agent $i$ is really faced with an $S_q$ picking sequence, and hence gets a value of at least $\omega(S_q, \items^i, v_i)$.

            \item In some future round $r$, the token jumps into $i$. Let $U$ denote the set of items that agent $i$ holds up to this round, and let $Y \subset \items^i$ be the items incident with $i$ that are not yet taken. Consider for $i$ the marginal valuation function $v_i^U$, satisfying $v_i^U(T) = v_i(T \cup U) \setminus v_i(U)$ for every set $T \subset Y$. Then, from the set $y$ of items, agent $i$ can guarantee $\omega(S_p, Y, v_i^U)$. By Proposition~\ref{pro:greater}, this is at least $\omega(S_q, Y, v_i^U)$. Summing over the whole game, the value that $i$ gets is at least $v_i(U) + \omega(S_p, Y, v_i^U) \ge v_i(U) + \omega(S_q, Y, v_i^U) \ge \omega(S_q, \items', v_i)$, as desired.
        \end{itemize}
        
    \end{enumerate}
    \end{proof}

    The following Corollary proves (among other things) Theorem~\ref{thm:graphical}.

    \begin{corollary}
        \label{cor:graphical}
    In allocation instances represented by multi-graphs as described above, if every agent $i$ has a subadditive valuation $v_i$ in which the maximum value of any single item is $\delta_i$, then there is an allocation in which every agent $i$ receives a bundle of value at least $\frac{v_i(\items) - \delta_i}{2}$. Moreover, the value received it at least $\frac{1}{2} MMS(\items, v_i, 2) \ge \frac{1}{2} MMS(\items, v_i, n)$.
    \end{corollary}

    \begin{proof}
        This is an immediate corollary of Theorem~\ref{thm:multigraph} and Corollary~\ref{cor:subadditive} (using the fact that $v_i(\items) = v_i(\items^i)$.
    \end{proof}

\subsection{From multi-allocations to allocations} 
\label{sec:induction}

We are now ready to prove Theorem~\ref{thm:main}. 




\begin{proof}
    Let $k$ be such that $d = 2^k$.

    We prove the theorem by induction on $k$. The base case, $k=1$, is Corollary~\ref{cor:graphical}. This is because every 2-multi-allocation $A_1, \ldots, A_n$ can be represented as a multi-graph instance in which $A_i$ is the set of edges incident with vertex $i$, and $v_i(A_i) = v_i(\items)$. 

    We now prove the induction step. Consider $k \ge 2$. We may assume that the theorem holds for $k-1$.   

    Make $2^{k-1}$ copies of each item $e$, denoted by $e_1, \ldots, e_{2^{k-1}}$. For each agent $i$ that holds $e$ in the $2^k$-multi-allocation $A = A_1, \ldots, A_n$, replace $e$ by one of the copies of $e$. There are sufficiently many copies so that each copy is given to at most two agents. Hence $A$ becomes a 2-multi-allocation with respect to the set of copies. By Corollary~\ref{cor:graphical}, it has an allocation $B$ in which for each agent $i$ we have $v_i(B_i) \ge \frac{v_i(A_i) - \delta_i}{2}$. 

    In $B$,  each item $e$ has at most $2^{k-1}$ copies. Hence $B$ is really is a $2^{k-1}$-multi-allocation. By the inductive hypothesis, there is an allocation $A'$, in which $v_i(A'_i) \ge \frac{v_i(B) - (2^{k-1} - 1)\delta_i}{2^{k-1}}$ for each agent $i$. Substituting into this expression the lower bound on $v(B_i)$, we get that $v_i(A'_i) \ge \frac{v_i(A_i) - (2^k - 1)\delta_i}{2^{k}}$, as desired.        
\end{proof}

\subsection{The subadditive MMS sampling lemma}
\label{sec:lemma}

In this section we prove Lemma~\ref{lem:sampling}, the subadditive MMS sampling lemma.

\begin{proof}
    Consider an arbitrary allocation instance $I$ with subadditive valuations. For every agent $i$, let $MMS_i$ denote the value of her MMS. For $d$ and $\rho$ as in the statement of the lemma, let $\alpha$ denote $\rho / (2\hat{d} - 1)$, the desired approximation for the MMS. If there is an item $e$ and an agent $i$ for which $v_i(e) \ge \alpha \cdot MMS_i$, then give item $e$ to agent $i$, and remove item $e$ and agent $i$ from the instance. All the removed agents get a value as required by the lemma. As to the remaining agents, let $I'$ denote the allocation instance that remains. The MMS of each agent in $I'$ (with respect to $n'$, the number of agents remaining in $I'$) is at least as large as it was in $I$ (it is well known that removing an agent and an item cannot decrease the MMS of the remaining agents).

    For $I'$, consider an $\rho$-MMS $d$-multi-allocation $A$ that is assumed to exist by the lemma. (Here, recall Remark~\ref{rem:sampling} that we may allow $d$ to be a monotone function of $n$. Monotonicity implies that $d(n') \le d(n)$, and so a $d(n')$-multi-allocation is also a $d(n)$-multi-allocation.) As $\hat{d} \ge d$, this allocation is also a $\hat{d}$-multi-allocation. As $\hat{d}$ is a power of~2, Theorem~\ref{thm:main} applies, and we have an allocation $A'$ in which each agent $i$ receives a value of at least $v_i(A'_i) \ge \left(\rho \cdot MMS_i  - \delta_i \cdot (\hat{d}-1)\right)/ \hat{d}$. In $I'$, for every agent $i$ it holds that $\delta_i < \rho \cdot MMS_i / (2\hat{d} - 1)$, and so $v_i(A'_i) \ge \rho / (2\hat{d} - 1)$, as desired.
\end{proof}

\subsection{Approximate MMS allocations}
\label{sec:loglogn}

In this Section we prove Corollary~\ref{cor:loglogn}. For this purpose, we use the sequence of integers $n_1 = 2$, $n_2 = 6$, $n_3 = 42$, $n_4 = 1806$, $\ldots$, defined by $n_1 = 2$ and $n_{d+1} = n_d(n_d + 1)$. 

A key lemma in our proof is the following result of~\cite{SS2025}.

\begin{lemma}[\cite{SS2025}]
    \label{lem:SS25}
    For every $d \ge 1$, every allocation instance with $n < n_d$ agents with subadditive valuations has a $\frac{1}{2}$-MMS $d$-multi-allocation.
\end{lemma}

For completeness, we present some background explaining how Lemma~\ref{lem:SS25} is derived.

The following result is known to be implied by results of~\cite{Feige09} (see~\cite{FG25}, for example).

\begin{lemma}[Provable using techniques of~\cite{Feige09}]
    \label{lem:feige09}
    In every allocation instance with subadditive valuations, there is a distribution over allocations in which every agent $i$:
    \begin{enumerate}
        \item Gets in expectation at least $\frac{1}{2}$-$MMS(\items, v_i, n)$.
        \item Gets value at least $\frac{1}{2}$-$MMS(\items, v_i, n)$ with probability at least $\frac{1}{2}$.
    \end{enumerate} 
\end{lemma}

Lemma~\ref{lem:feige09} was extended by~\cite{SS2025} in an insightful way to situations in which the benchmark of $MMS(\items, v_i, n)$ is replaced by $MMS(\items, v_i, k n)$, for any positive integer $k$. Though this weakens the first item in the lemma, it allows for improved probability in the second item. 

\begin{lemma}[\cite{SS2025}]
    \label{lem:SS25a}
    For every integer $k \ge 1$, in every allocation instance with subadditive valuations, there is a distribution over allocations in which every agent $i$:
    \begin{enumerate}
        \item Gets in expectation at least $\frac{1}{2}$-$MMS(\items, v_i, k n)$.
        \item Gets value at least $\frac{1}{2}$-$MMS(\items, v_i, k n)$ with probability at least $\frac{k}{k+1}$.
    \end{enumerate}
\end{lemma}


A consequence of item~2 of Lemma~\ref{lem:SS25} is that there is an allocation that gives at least $\frac{kn}{k+1}$ agents their respective  $\frac{1}{2}$-$MMS(\items, v_i, k n)$ value.

Applying this consequence repeatedly on agents that remain (each time with the whole set $\items$ of items), one gets that for every positive integer $d$, there is a $d$-multi-allocation in which at least $\frac{(n_d - 1)n}{n_d}$ agents get their respective  $\frac{1}{2}$-$MMS(\items, v_i, n)$ value. Hence, for $n < n_d$, all agents get their respective  $\frac{1}{2}$-$MMS(\items, v_i, n)$ value, proving Lemma~\ref{lem:SS25}.

Observe that for every $d \ge 2$, it holds that $n_d > 2^{2^{d-1}}$, implying that $d < \log\log n_d + 1$. Thus, the combination of Lemma~\ref{lem:SS25} and Lemma~\ref{lem:sampling} implies Corollary~\ref{cor:loglogn}.


It is instructive to see what Corollary~\ref{cor:loglogn} implies for specific values of $d$. As Lemma~\ref{lem:sampling} rounds $d$ up to the nearest power of~2, we need only consider values of $d$ that are a power of~2.

For $d=2$ we have $n_d = 6$, implying that $\frac{1}{6}$-MMS allocations exist for all $n \le 5$. Stronger results are known, such as $\frac{1}{n}$-MMS allocations for all $n$~\cite{feige2025residualmaximinshare}, and Proposition~\ref{pro:n8}.

For $d=4$ we have $n_d = 1806$, implying that $\frac{1}{14}$-MMS allocations exist for all $n \le 1805$. As a function of $m$, it was known that $\frac{1}{14}$-MMS allocations exist for all $m \le 13^{13}$~\cite{FG25}.

For $d=8$ we have $n_d > 1806^{16} > 10^{50}$, implying that $\frac{1}{30}$-MMS allocations exist for all $n \le 10^{50}$.

It is difficult to imagine allocation instances with more than $10^{50}$ agents, so we will stop here.

For every $n \ge 4$ (a technical condition that ensures that $\log\log n \ge 1$), the best of $\frac{1}{n}$ and the bounds listed above (extending the above table to values of $d$ that are higher powers of two, and for each $n$, taking the appropriate $n_d$) are at least $\frac{1}{8\log\log n}$.

\section{Extensions}

For a vector ${\bf d} = (d_1, \ldots, d_n)$ of positive integers, a ${\bf d}$-multi-allocation is one in which for every agent $i$, every item that is allocated to agent $i$ is allocated to at most $d_i$ agents (including $i$). The proof of Theorem~\ref{thm:main} extends to show the following. 

\begin{theorem}
    \label{thm:extension}
    Let ${\bf d} = (d_1, \ldots, d_n)$ be such every $d_i$ is a perfect power of~2. Let $A = A_1, \ldots, A_n$ be a $\bf{d}$-multi-allocation. Suppose every agent $i$ has a subadditive valuation $v_i$ with $\max_{e\in A_i} v_i(e) = \delta_i$. Then there exists an allocation $A' = A'_1, \ldots, A'_n$ in which 
    $$v_i(A'_i) \ge \frac{v_i(A_i)}{d_i} - \frac{d_i-1}{d_i}\delta_i$$
    for every agent $i$.
\end{theorem}

We now consider instances in which agents have arbitrary, perhaps unequal, entitlements. In this setting, every agent $i$ has an entitlement $b_i > 0$, with $\sum b_i = 1$. (In the case of equal entitlement, $b_i = \frac{1}{n}$ for all $i$.) The fairness notion that we use in this case is the anyprice share (APS)~\cite{BEF21APS}.

\begin{definition}
    For a set $\items$ of items and a fraction $\rho$, a fractional $\rho$ partition is a collection of bundles $\{P_j\}$ (where $j$ ranges over some index set) and associated nonnegative weights $\{\lambda_j\}$ satisfying:

    \begin{itemize}
        \item $\sum_j \lambda_j = 1$.
        \item $\sum_{j \; \mid \; e \in P_j} \lambda_j = \rho$ for every item $e\in \items$.
    \end{itemize}

    The set of all fractional $\rho$ partitions is denoted by ${\cal{FP}}_{\rho}$.

    For an agent $i$ with valuation $v_i$ and entitlement $b_i$, her anyprice share (APS) is:

    $$APS(\items,v_i,b_i) = \max_{(P_1, P_2, \ldots) \in {\cal{FP}}_{b_i}} \min_j v_i(P_j)$$
\end{definition}


\begin{theorem}
        \label{thm:graphicalAPS}
    Graphical allocation instances with agents with arbitrary entitlements and subadditive valuations have $\frac{1}{2}$-$APS$ allocations. 
    \end{theorem}

The proof is similar to that of Theorem~\ref{thm:graphical}, and is omitted. 

\begin{remark}
 Theorem~\ref{thm:graphicalAPS} can be further strengthened to show the existence of $\frac{1}{2}$-$\widehat{APS}$ allocations. $\widehat{APS}$ is similar to APS, except that entitlements are rounded up to the nearest inverse integer (this operation was studied in the context of MMS in \cite{BF24}). That is, if $\frac{1}{k+1} < b_i \le \frac{1}{k}$ for integer $k$, then $\widehat{APS}(\items, v_i, b_i) = APS(\items, v_i, \frac{1}{k})$. In particular, if $\frac{1}{2} < b_i < 1$, then $\widehat{APS}(\items, v_i, b_i) = v_i(\items)$. There can be at most one agent with $b_i > \frac{1}{2}$. For this agent, the guarantee of $\omega(S_q,\items^{i},v_i)$ in the statement of Theorem~\ref{thm:multigraph} does not suffice. However, in the proof of Theorem~\ref{thm:multigraph}, one agent of our choice (the first agent to hold the token) gets the stronger guarantee of $\omega(S_p,\items^{i},v_i)$, which by Corollary~\ref{cor:subadditive} is at least  $\frac{1}{2}v_i(\items)$, as desired.
\end{remark}


\section{Discussion}

We leave open the question of whether Theorem~\ref{thm:main} holds when $d$ is not a power of~2. If it does, then in the subadditive sampling lemma (Lemma~\ref{lem:sampling}) the term $\hat{d}$ can be replaced by $d$.

Lemma~\ref{lem:sampling} may have future application to approximate MMS allocations, if $d$-multi-allocations with smaller values of $d$ will be discovered. We note that for $d < n$, there are examples showing that $\rho$-MMS $d$-multi-allocations with $\rho > \frac{1}{2}$ do not exist. Hence, the most optimistic set of parameters is $\frac{1}{2}$-MMS $2$-multi-allocations. If they can be shown to exist, then Lemma~\ref{lem:sampling} would imply the existence of $\frac{1}{6}$-MMS allocations. Recall that the upper bound known on $\rho_{SA}$ is $\frac{1}{2}$, so it seems that other techniques are required if one is to determine the exact value of $\rho_{SA}$.

\subsection*{Acknowledgments}

This research was supported in part by the Israel Science Foundation (grant No. 1122/22).

\bibliographystyle{alpha}


\newcommand{\etalchar}[1]{$^{#1}$}

\end{document}